\documentclass[fleqn,twoside]{article}
\usepackage{espcrc2}

\usepackage{graphicx}

\title{
  Exploration of sea quark effects in two-flavor QCD with the
  $O(a)$-improved Wilson quark action\thanks{
    Talk presented by S.~Hashimoto}
  }

\author{
  JLQCD Collaboration:
  S.~Aoki
  \address{
    Institute of Physics, University of Tsukuba, 
    Tsukuba, Ibaraki 305-8571, Japan
    },
  R.~Burkhalter$^{\rm a,}$
  \hspace{-1mm}
  \address{
    Center for Computational Physics, 
    University of Tsukuba, Tsukuba, Ibaraki 305-8577, Japan
    }, 
  M.~Fukugita
  \address{
    Institute for Cosmic Ray Research,
    University of Tokyo, Kashiwa, Chiba 277-8582, Japan
    }, 
  S.~Hashimoto
  \address{
    High Energy Accelerator Research Organization (KEK), 
    Tsukuba, Ibaraki 305-0801, Japan
    },
  K-I.~Ishikawa$^{\rm d}$,
  N.~Ishizuka$^{\rm a,b}$, 
  Y.~Iwasaki$^{\rm a,b}$,
  K.~Kanaya$^{\rm a,b}$, 
  T.~Kaneko$^{\rm d}$,
  Y.~Kuramashi$^{\rm d}$, 
  M.~Okawa$^{\rm d}$,       
  T.~Onogi
  \address{
    Yukawa Institute for Theoretical Physics,
    Kyoto University, Kyoto 606-8502, Japan
    }, 
  S.~Tominaga$^{\rm b}$,
  N.~Tsutsui$^{\rm d}$,
  A.~Ukawa$^{\rm a,b}$,
  N.~Yamada$^{\rm d}$,
  and 
  T.~Yoshi\'e$^{\rm a,b}$.
} 

\begin{document}

\begin{abstract}
  We explore sea quark effects in the light hadron mass spectrum in
  a simulation of two-flavor QCD using the nonperturbatively
  $O(a)$-improved Wilson fermion action.
  In order to identify finite-size effects, light meson masses are
  measured on $12^3\times 48$, $16^3\times 48$ and $20^3\times 48$
  lattices with $a\sim$ 0.1~fm.
  On the largest lattice, where the finite-size effect is negligible,
  we find a significant increase of the strange vector meson mass
  compared to the quenched approximation. 
  We also investigate the quark mass dependence of pseudoscalar meson
  masses and decay constants and test the consistency with 
  (partially quenched) chiral perturbation theory.
\end{abstract}

\maketitle

\section{Introduction}

While encouraging results are being accumulated on the effect of
dynamical quarks in the light hadron spectrum
\cite{Aoki:2001kp}, 
further effort is needed to establish its presence and magnitude with
precision.  
In this study we explore the sea quark effects in two-flavor QCD with
the $O(a)$-improved Wilson fermion.
In particular, we analyze the strange vector meson masses in detail, 
for which quenched QCD fails to reproduce the experimental values 
\cite{Aoki:2000yr}.
Since the hadron spectrum may be distorted by a finite box size 
especially for light sea quark masses, we carry out simulations with
three volumes in order to identify the finite size effect. 

Another point of our study is a test of chiral perturbation theory for 
pseudoscalar meson masses and decay constants. 
The appearance of the chiral logarithm in the quark mass, which
depends only on the number of active flavors, provides a definite test
of the sea quark effect.

\section{Two-flavor QCD simulations}

The JLQCD collaboration has been performing a two-flavor QCD
simulation with the standard glue and the $O(a)$-improved Wilson quark
actions at $\beta=5.2$ \cite{Aoki:2001yi}.
The improvement coefficient $c_{\mathrm{sw}}$ is determined
nonperturbatively \cite{Jansen:1998mx} as $c_{\mathrm{sw}}$=2.02, 
which we confirmed in the course of this study. 
We choose five values of the hopping parameter $\kappa$ ($\kappa$ =
0.1340, 0.1343, 0.1346, 0.1350, and 0.1355) to cover the range
$m_{PS}/m_V$ = 0.8--0.6.
The effective lattice spacing determined through the static quark
potential changes from $a^{-1}$ = 1.6~GeV to 2.0~GeV as the sea quark
mass decreases.

In order to detect finite size effects we perform simulations on
three lattices $12^3\times 48$, $16^3\times 48$, and
$20^3\times 48$.
The simulations run through 3000 HMC trajectories for each $\kappa$.
An additional 3000 trajectories is performed with a better
preconditioning \cite{Ishikawa_lat2001} for the $20^3$ lattice.
Some earlier results on the $16^3$ lattice were already
presented at the last lattice conference \cite{Aoki:2001yi}.

\section{Vector meson mass}

\begin{figure}[t]
  \begin{center}
    \leavevmode
    \includegraphics*[width=6.9cm,clip]{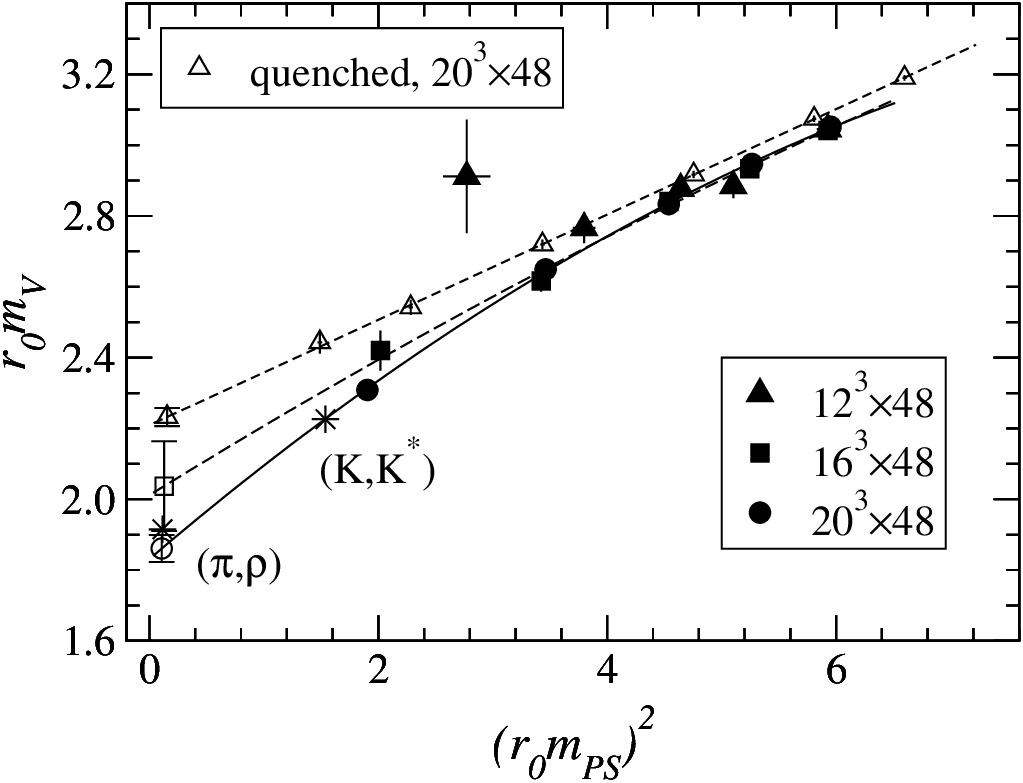}
  \end{center}
  \vspace*{-12mm}
  \caption{
    Degenerate vector meson masses in quenched and two-flavor QCD. 
    Stars represent the physical points with $r_0$=0.49~fm.
    }
  \vspace*{-4mm}
  \label{fig:mrho_vs_mpi2}
\end{figure}

We plot results for degenerate pseudoscalar and vector meson masses 
on three lattice volumes in Figure~\ref{fig:mrho_vs_mpi2} together
with quenched data. 
Finite-size effects badly affect the masses on the smallest volume 
($12^3$) except for the heaviest point.
The next largest lattice ($16^3$) suffers less, but we still find 
significant difference from the largest lattice ($20^3$) at the
lightest data point, which corresponds to $m_{PS}/m_V \simeq$ 0.6.
The effect is about 3\% (1.9 $\sigma$) for pseudoscalar and 5\% (2.0
$\sigma$) for vector mesons.
Our observation implies that the effective lattice extent (at finite
sea quark mass) should be kept larger than 2~fm in order to avoid
finite size effects in (lowest lying) light meson masses for 
$m_{PS}/m_V \geq$ 0.6--0.7.
A similar study indicates that the corresponding spatial extent for 
baryons is at least 2.5~fm. 

In Figure~\ref{fig:mrho_vs_mpi2} we normalize masses by $r_0$ to 
absorb the change of effective lattice spacing for different sea 
quark masses.  
We find a significant curvature for the $20^3$ lattice, in contrast to
the quenched case for which no signal of curvature is observed.
By a quadratic fit to reach the chiral limit we obtain 
$r_0 m_\rho$ = 1.86(4), consistent with the phenomenological
value $r_0$ = 0.49~fm.

Strange meson masses are obtained by measuring non-degenerate mesons
for which sea and valence quark masses are different.
Taking the chiral limit for sea quark we obtain $K$-$K^*$-like and
$\eta$-$\phi$-like combinations. 
Using the $K$ meson mass as input for the strange quark mass we obtain 
$m_{K^*}/m_\rho$ = 1.15(1) and $m_\phi/m_\rho$ = 1.29(2). 
These results are significantly higher than the quenched results and
closer to the experimental values.

\begin{figure}[t]
  \begin{center}
    \leavevmode
    \includegraphics*[width=6.9cm,clip]{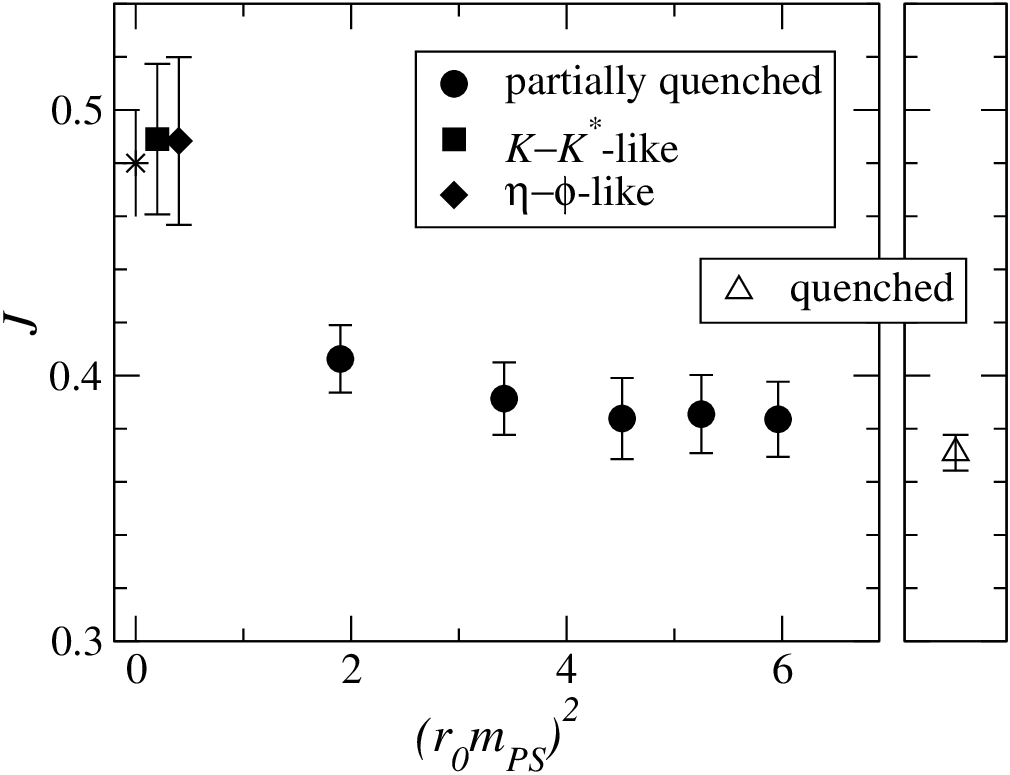}
  \end{center}
  \vspace*{-12mm}
  \caption{
    $J$ parameter as a function of sea quark mass.
    A quenched result is in the right panel.
    }
  \vspace*{-4mm}
  \label{fig:J}
\end{figure}

The $J$ parameter \cite{Lacock:1995tq} may be calculated for a fixed
sea quark mass varying valence quark masses.
The results are plotted in Figure~\ref{fig:J} as a function of 
$(r_0 m_{PS})^2$.
Compared to the quenched estimate, the two-flavor value of $J$ is 
slightly higher and shows a trend of increase toward the chiral
limit. 
The $J$ parameter can also be constructed from the $K$-$K^*$-like and 
$\eta$-$\phi$-like combinations.
These are consistent with the phenomenological estimate as shown in
Figure~\ref{fig:J}.

\section{Test of chiral perturbation theory}

In order to obtain a controlled chiral limit of the lattice data, 
chiral perturbation theory (ChPT) may be used as a guide. 
Prior to adopting this strategy, it is worth testing if lattice data 
are consistent with the chiral behavior predicted by ChPT,
especially the chiral logarithms.

For $N_f$ flavors of degenerate quarks with a mass $m_S$, 
the pseudoscalar meson mass $M_{\mathrm{SS}}$ to one-loop order of
ChPT is \cite{Gasser_Leutwyler} given by
\begin{eqnarray}
  \label{eq:ChPT_PSmass}
  \lefteqn{
    \frac{M_{\mathrm{SS}}^2}{2B_0 m_S}
    = 1 + \frac{1}{N_f} y_{\mathrm{SS}}\ln y_{\mathrm{SS}}
    } \nonumber\\
  & & 
  + y_{\mathrm{SS}} [
  (2\alpha_8-\alpha_5) + N_f (2\alpha_6-\alpha_4)
  ]
\end{eqnarray}
with $y_{\mathrm{SS}}=2B_0m_S/(4\pi f)^2$.  
While the low energy constants $\alpha_i$ are unknown parameters,
the chiral log term $y_{\mathrm{SS}}\ln y_{\mathrm{SS}}$ appears
with a definite coefficient depending only on the number of flavors.

\begin{figure}[tb]
  \begin{center}
    \leavevmode
    \includegraphics*[width=6.9cm,clip]{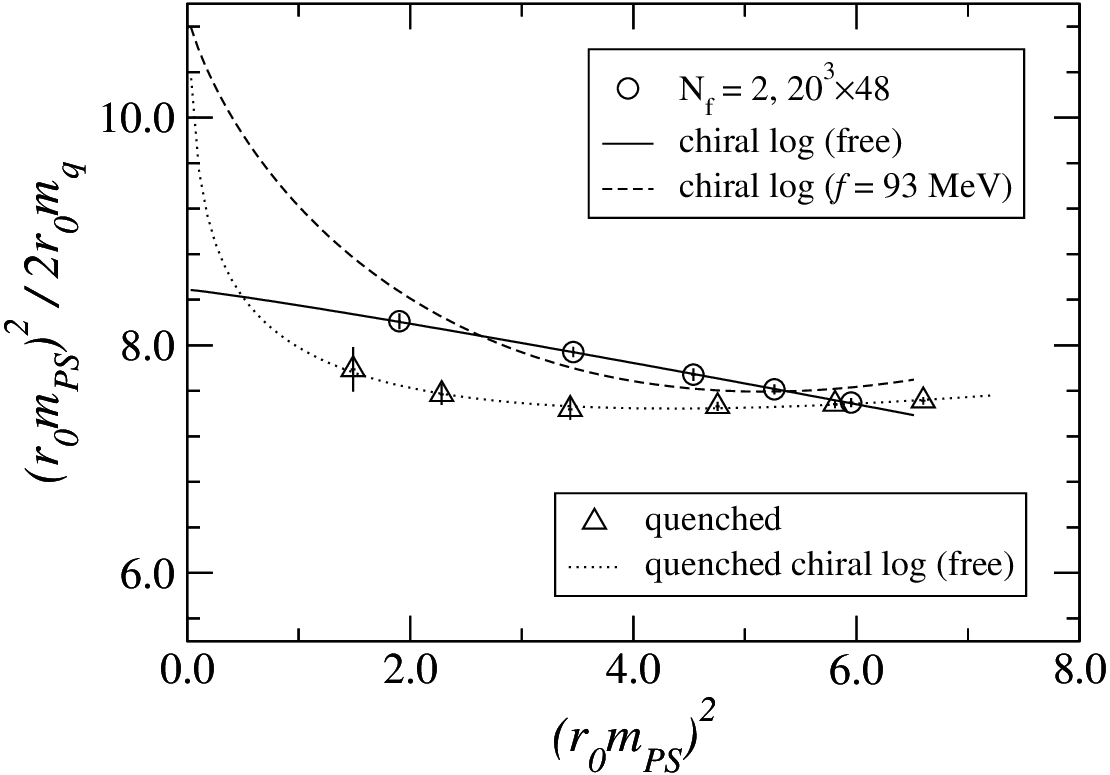}
  \end{center}
  \vspace*{-12mm}
  \caption{
    Test of the ChPT formula for the pseudoscalar meson mass.
    }
  \vspace*{-4mm}
  \label{fig:B_vs_mpi2}
\end{figure}

Figure~\ref{fig:B_vs_mpi2} shows the unquenched and quenched
results for $M_{\mathrm{SS}}^2/2m_S$.
While there is a visible difference between the two sets of results, 
the two-flavor data do not show the curvature expected from 
the chiral logarithm shown by a dashed curve.
The situation is the same for the chiral behavior of the pseudoscalar
decay constant.

We also perform a test for non-degenerate mesons using the partially 
quenched ChPT (PQChPT) \cite{Sharpe:1997by} by considering a ratio for
which the low energy constants cancel 
\begin{equation}
  \label{eq:ratio_tests}
  \frac{
    \left( \frac{M_{\mathrm{VS}}^2}{m_V+m_S} \right)^2
      }{
      \frac{M_{\mathrm{SS}}^2}{2m_S}
      \cdot
      \frac{M_{\mathrm{VV}}^2}{2m_V}
      }
    =
    1 + \frac{y_{\mathrm{SS}}}{N_f} t,
\end{equation}
where 
$t\equiv \ln \frac{y_{\mathrm{VV}}}{y_{\mathrm{SS}}}
+ 1 - \frac{y_{\mathrm{VV}}}{y_{\mathrm{SS}}}$.
The subscript $V$ denotes a valence quark whose mass may be different
from the sea quark mass.
In our preliminary results we find that the coefficient of 
$t$, the chiral logarithm term, is much smaller than expected.
A similar result is also observed for the decay constants.

A possible reason for the deficit of the chiral logarithm is that the
sea quark mass in our simulations is still too large to be described
by ChPT.
In fact, if one introduces a singlet meson $\eta'$ into PQChPT
\cite{Golterman:1998st} the chiral logarithm is substantially
suppressed unless $M_{\mathrm{SS}}^2$ is much smaller than $m_0^2/3$,
the additional mass for the singlet meson.
In our simulation, however, $M_{\mathrm{SS}}^2/(m_0^2/3)$ is $O(1)$.

\section{Conclusions}

In the two-flavor QCD simulation we find an evidence of the sea
quark effect in the strange vector meson masses.
A large enough lattice volume, at least (2 fm)$^3$, is necessary
to identify the sea quark effect without suffering from the finite
size effect.
On the other hand, the lattice data for the pseudoscalar meson
masses and decay constants fail to reproduce the chiral
logarithms, suggesting that the sea quark mass corresponding to
$m_{PS}/m_V\ge$~0.6 is still too large to be described by ChPT.
We are currently accumulating further statistics to confirm these
findings.

\vspace{6mm}
\noindent
Numerical simulations are performed on Hitachi SR8000 model F1
supercomputer at KEK under the Supercomputer project No.~66 (FY 2001).
This work is also supported in part by the Grant-in-Aid of the
Ministry of Education (Nos.~10640246, 11640294, 12014202, 12640253,
12640279, 12740133, 13640260, and 13740169).
K-I.I. and N.Y. are supported by JSPS.

\end{document}